# DAMA/LIBRA FINDINGS URGE REPLACEMENT OF THE WIMP HYPOTHESES BY THE DAEMON PARADIGM AS A BASIS FOR EXPERIMENTAL STUDIES OF DM OBJECTS

E.M.Drobyshevski

*Ioffe Physical-Technical Institute, Russian Academy of Sciences, 194021 St-Petersburg, Russia*
*emdrob@mail.ioffe.ru*

The simplest version of the daemon paradigm suggests the modulated 2-6-keV range events in DAMA/NaI and DAMA/LIBRA detectors are caused by the iodine ions knocked out elastically by the electrically neutral c-daemons moving with $V$ = 30-50 km/s (c-daemon is a complex of negative daemon located in a remainder of formerly captured nucleus where the daemon decomposes nucleons one by one with ~$10^{-6}$ s mean interval). Furthermore, after the 2-6 keV event occurred, in subsequent ~$10^{-6}$ s, the c-daemon (which becomes negative during this time) recaptures new nucleus with resulting scintillations in ~10 MeV range! The last possibility was so far overlooked in the experiments as it did not stem from WIMP hypotheses. A modification of the NaI(Tl) experiments is suggested for revealing the effect described. Independently of the outcome, any obtained result will be important for refining the daemon paradigm further on.

*The running title*: Daemons in NaI(Tl): keV events are ~1 μs precursors of the MeV ones

## 1. Introduction

The DAMA/NaI and DAMA/LIBRA systems, with many 3×3 and 5×5 NaI(Tl) crystals, respectively, each measuring 10.2×10.2×25.4 cm³ and 9.7 kg in mass, were intended from the very beginning to detect signals which were expected to be generated by the recoil nuclei knocked out elastically in the crystal by WIMPs (Weakly Interacting Massive Particles), i.e., DM objects of the galactic halo moving with $V$ ~ 200-300 km/s [1,2]. Assuming a WIMP mass of ~100 GeV, the kinetic energy $E_r$ of the recoil nuclei of iodine should amount to tens of keV, and that of Na nuclei, accordingly, about one fifth of that (it is for such nuclear recoil energies $E_r \geq 10$ keV that the detectors based on Si, Ge and other such crystals are designed). Actually, the DAMA systems detected signals of substantially lower energies, in the $E_r$ = 2-6-keV range of electron equivalent, which were found to be modulated with a period $P$ = 1 y [1,2]. It turns out that this range matches nicely with the velocities of supermassive particles (e.g., daemons; see Sec. 2 below) with $V$ = 30-50 km/s which fall from strongly elongated Earth-crossing heliocentric orbits (SEECHOs), their flux, as estimated from DAMA/NaI data (with due account of the recoil ion channeling), being $f$ ~ 6×$10^{-7}$ cm$^{-2}$s$^{-1}$ [3]. This value is in accord with our earlier estimates [4] and is not in conflict with measurements of the daemon flux from near-Earth, almost circular heliocentric orbits (NEACHOs) ($f > 10^{-7}$ cm$^{-2}$s$^{-1}$) [5,6].

The crucial point in DAMA measurements was application of the so-called single-hit criterion (SHIC), a scenario by which the first event with $E_r > E_0$ = 2 keV in one scintillator vetoes 600 ns later recording of signals from any scintillator during the following 500 μs (which introduces a systematic error into the measured rate of about $10^{-4}$ [1], whence, by the way, we obtain immediately ~5 s for the average interval between successive triggers of the system). This approach may be justified for WIMPs which traverse in condensed matter without interaction with nucleons a distance measured in light years (the cross section of interaction $\sigma$ ~ $10^{-43}$-$10^{-42}$ cm²/nucleon). Now if, however, the DM particles are not WIMPs but, say, daemons (with a large gas-dynamic cross section $\sigma$ ~ $10^{-19}$ cm² [7]), one should invoke the SHIC with extreme care, otherwise the potential inherent in a massive detector will not be used to its full capacity, the conclusion suggested by the operation of DAMA/LIBRA; indeed, the number of 2-6-keV events per 1 kg of mass recorded with this equipment and featuring modulation with $P = 1$ y has turned out to be close to one half only of the DAMA/NaI events (thus verifying our expectations [8,9]).

We are going to show now that a certain modification of the DAMA NaI(Tl) detectors could not only lend support to the daemon paradigm but provide essential information on many properties of the daemons as well.

## 2. Essentials of the daemon paradigm and its experimental verification

In planning our experiments on detection of negative daemons (DArk Electric Matter Objects, tentatively assumed to be Planckian black holes with $M \sim 10^{-5}$ g carrying an electric charge corresponding to their mass, $Ze \approx 10e$) which build up in NEACHOs and strike the Earth with $V = 11.2$-$15$ km/s, we started from somewhat different prerequisites.

It was assumed that in traversing a material, a multiply negatively charged daemon is capable of capturing an atomic nucleus with a binding energy $W \approx 1.8ZZ_nA_n^{-1/3}$ MeV (for I it is 192 MeV, for Zn and Cu - 134 MeV, for F, Ne, Na - 61, 66 and 70 MeV, respectively) [10,11]. Liberation of such an energy shakes off the atomic and refilling electrons and initiates emission from the nucleus (during $\sim 10^{-9}$-$10^{-8}$ s [11]) of nucleons, their complexes and $\gamma$ rays, a process that can generate a strong scintillation. The daemon that has captured a nucleus with $Z_n \geq 24/Z$ turns up inside it (a situation familiar from muon physics for heavy enough nuclei) and, apparently, can even penetrate into a proton. The daemon becomes poisoned, as it were, for a certain time by the remainder of the heavy nucleus (we shall refer to this complex as the c-daemon), so that, if the charge of the latter is larger than $Z$, the c-daemon will interact with matter as a conventional atom (we keep in mind, however, that, having a giant mass and the corresponding kinetic energy, it should traverse almost freely the material by inertia). Next, we ventured the simplest and fairly natural assumption that on entering a nucleon (proton) the daemon will catalyze its decay in $\Delta\tau_{ex} \sim 10^{-6}$ s [11], with the result that a certain time later the c-daemon will acquire a zero, and, subsequently, again the negative electric charge. A daemon with $Z_{eff} = -1$ ($<0$) and $V = 10$-$30$ km/s is capable of capturing in condensed matter another nucleus after travelling a distance $\sim 1$-$10$ μm [11], accompanied by another release of energy $\sim 10^2$ MeV and, apparently, the loss of the remainder of the preceding nucleus.

This ideology underlies development of our detector with two thin ($\sim 10$ μm) ZnS(Ag) scintillators (or with one such ZnS(Ag) scintillator and a FEU-167 PM tube with a thick internal Al coating serving as a second sensor [5,6]) which detects reliably (by now, C.L. > 99.99%) a flux of NEACHO daemons $f > 10^{-7}$ cm$^{-2}$s$^{-1}$ varying with $P = 0.5$ y (with maxima in March and September) [5,6]. The behavior of this stream, just as of the stream of particles with $V = 30$-$50$ km/s with $P = 1$ y, finds a straightforward explanation within the concepts of celestial mechanics [7], provided one admits the capture of galactic disc daemons into SEECHOs and, subsequently, into NEACHOs [4-8].

Our observations lend apparently credence to the above scenario and yield $\Delta\tau_{ex} \sim 10^{-6}$ s for the daemon-stimulated proton decay time [11]. {We note, however, that we have not yet succeeded in directly demonstrating the decay itself of nucleons (protons) with a release of energy close to 938 MeV in experiments with a thick (4.3 cm) CsI(Tl) scintillator [12] (the modes of the daemon-stimulated decay can certainly be rather nontrivial, including, possibly, emission of gravitons etc.). We may add that DAMA (just as the other systems, apparently designed particularly for this purpose, like Kamiokande etc.) likewise would not record daemon-stimulated proton decay, while it responds, in a certain sense, to a product of these events, more specifically, to variable-charge c-daemons.} Therefore, we are going to entertain the above simplest scenario, because it appears to find support both from our and the DAMA experiments. We shall use it as a basis to propose scenarios of somewhat modified experiments of the type of DAMA/NaI and DAMA/LIBRA with massive scintillator assemblies.

## 3. 2-6-keV signals as precursors of stronger, MeV-range, events in the NaI(Tl) detector

In the frame of the simplest version of the daemon paradigm, the 2-6-keV signals detected in the DAMA NaI(Tl) scintillators are generated by iodine ions knocked out elastically by neutral c-daemons. A c-daemon resides in an electrically neutral state, on the average, for $\Delta\tau_{ex} \sim 10^{-6}$ s, until the next daemon-stimulated proton decay occurs in the remainder of the nucleus captured by the daemon [11]. When this happens, the daemon acquires a negative charge $Z_{eff}= -1$ (<0) for the subsequent ~$10^{-6}$ s. While the c-daemon carrying zero charge traverses with $V$ = 30-50 km/s a distance in matter, on the average, $V \times \Delta\tau_{ex} \approx$ 3-5 cm, which practically coincides with its path in NaI between two elastic collisions with atomic nuclei (with an elastic scattering cross section $\sigma$ = 5-10 barn, a figure characteristic of neutrons), the path crossed by a daemon with $Z_{eff} = -1$ (<0) before capture of the next nucleus is measured in a few tens of μm only [11]. The energy released in the capture of a nucleus is ~100 MeV. This scenario turned out fruitful in development of our NEACHO daemon detector consisting of thin ZnS(Ag) scintillators. It is the high-energy capture of the new nucleus in the thin ZnS(Ag) scintillator (and/or in the bottom lid of the tin-iron sheet casing or in the PM tube itself with a thickened internal Al coating) that generates our signals with a MeV-scale energy equivalent.

Application of SHIC in DAMA/NaI and DAMA/LIBRA, a scenario by which 600 ns after the arrival of the first signal with $E_r > E_0 = 2$ keV the electronics vetoes for 500 μs all the subsequent signals, is seen to automatically exclude detection of a much stronger, but coming later in time, MeV-range signal which was initiated by the (re)capture of a new nucleus made by the negative c-daemon. It is difficult to say to what extent the selection of the 600-ns time window, with some multiple-hits events (with $E \le 2.5$ MeV, see Sec. 4 below) recorded too, was made accidentally. In any case, annual modulation of the number of multiple-hits events is not seen with this window, thus suggesting that this modulation in the 2-6-keV range does not originate from some local background factors indeed [1,2].

In our previous publication [9] we ventured a suggestion that it would be appropriate to extend the time window in LIBRA to a few tens of μs in order to be able to detect movement of particles with $V$ = 30-50 km/s. The significance of this modification can be illustrated as follows. Capture of an iodine nucleus by a c-daemon with $Z_{eff} = -1$ (< 0), which was embedded in the nuclear remainder, a fluorine-type nucleus, releases about 130 MeV ($W_I - W_F \approx$ 192 - 61 = 131 MeV, whereas capture of a Na nucleus would liberate 9 MeV only), an energy large enough to evaporate ~12-15 nucleons (and of their complexes), which are moving with ~1-3 MeV energy each, and transmute the captured iodine nucleus, for instance, to the In nucleus ($Z_n = 49$, $A_n = 113$). Subsequent daemon-stimulated proton decays create in $\Delta\tau_{ex} \times (Z_n - Z) \approx 39$ μs a c-daemon with $Z_{eff} = 0$, which is now also capable of knocking out elastically an iodine nucleus with $E_r$ = 2-6 keV. During this time, a c-daemon moving with $V$ = 30-50 km/s traverses a distance of 1-2 m, which is comparable with the dimensions of DAMA/LIBRA, and furthermore, with the size of the 1-t NaI(Tl) system being planned by the DAMA collaboration.

In view of the above, however, we see that the 2-6-keV "elastic" signals may possibly be ~1-μs "precursors" of much stronger, MeV-range (up to ~10 MeV) signals caused by the capture of a new nucleus by the c-daemon, which, on having digested one more proton, lowers its charge to a negative value. It would therefore be extremely interesting to have a provision of detecting double-hit events within a time not of 0.6 μs but, say, of 5 or even 10 μs (indeed, the accuracy with which we know $\Delta\tau_{ex}$ is not high; the same applies to the cross section of elastic interaction of neutral c-daemons with atomic nuclei). One cannot overlook the possibility that the 2-6-keV events which do exhibit the $P$ = 1 y modulation can be followed $\ge 1$ μs later by MeV-range scintillations excited mainly in neighboring NaI(Tl) crystals.

Interestingly, capture of a Na nucleus entailing liberation of as little as about 9 MeV, an energy sufficient for evaporation of one nucleon only or emission of several γ-quanta, after which, 1-2 μs later, the c-daemon becomes neutral again for the subsequent $\Delta\tau_{ex} \sim 10^{-6}$ s, can also be treated as a precursor of a probable next multi-MeV capture of a much heavier nucleus of iodine.

Should such keV- and MeV-range precursor events be detected, their analysis would offer a possibility of refining the (averaged) value of $\Delta\tau_{ex}$, let alone many other ramifications, including naturally substantiation of the daemon paradigm itself, with the attendant determination of a possible value of the daemon charge $Z$ (indeed, it is possible that $Z$ = -9 rather than -10 [13]).

In the case of a positive result (i.e., confirmation that the 2-6-keV events are indeed precursors) we shall be left with wonder about how the DAMA experimenters managed to choose a window of close to optimal value of 600 ns, slightly less than $\Delta\tau_{ex} \sim 10^{-6}$s (again an intriguing coincidence), which permitted them to utilize to the utmost the unique potential of their NaI(Tl) crystals and separate the events associated with low-energy elastic collisions of neutral c-daemons with atomic nuclei from the much stronger capture-related multi-MeV events.

## 4. "Elephan'? Nope, Sir, can't say I saw 'im", or why the DAMA NaI(Tl) systems do not detect the MeV-range events of daemon capture of nuclei?

The headline of this Section refers to the fable written in 1814 by Russian poet Ivan Krylov about a man who came to the Museum of Natural History, was fascinated by picturesque butterflies and other tiny creatures on display and left overwhelmed with emotions, without noticing the staffed elephant towering over the visitors of the Museum.

Indeed, why the high-sensitivity DAMA/NaI and LIBRA experiments have not yet detected daemon capture involving MeV-range release of energy? The answer to this question is fairly straightforward: "One will not find what one is not looking for" [14]. The electronics of these experiments was tuned to detection of single events (with the SHIC, see Sec. 1 above), with an energy release of 2-2500 keV [1,2,15,16] (the latter value comes from calibration with $^{60}$Co) and even up to 10-36 MeV where no event was observed at all [17]. The situation is quite understandable as any multi-MeV process is accompanied by γ ray emissions causing scintillations in neighboring crystals which registration is forbidden by the SHIC but again. Such a turn could be hard to expect while remaining in the frame of the WIMP concept; indeed, comprehensive studies of the response of the system and of the pattern of detected events were performed up to $E \approx 90$ keV [1,16], well above the levels expected within the WIMP hypothesis.

One could have put a full stop here, were it not for certain attendant aspects.
(1) The separation of scintillation signals from noise, judging by their oscilloscopic trace area, was based on two criteria, $X$ and $Y$ by [1,15], where $X$ = (Area from 100 ns to 600 ns)/(Area from 0 ns to 600 ns) and $Y$ = (Area from 0 ns to 50 ns)/(Area from 0 ns to 100 ns). The scintillation signals crowd in the region of $X$ = 0.7 and $Y$ = 0.5, whereas for noise signals $X \to 0$, $Y \to 1$. If the beginning of the MeV-range signal falls within the 600-ns window after the system has been triggered by a 2-6-keV signal, we clearly obtain $X \to 1$, $Y$ is not definite, i.e., this signal does not fall into the region of possible WIMP events in the $XY$ coordinates and, thus, should be dropped.
(2) The time interval between the 2-6-keV signal initiated by an iodine ion knocked elastically out by a neutral c-daemon (an event that may occur at any instant as long as the c-daemon stays in the NaI(Tl) crystal) and the MeV-range signal caused by the capture of a nucleus by the same c-daemon but which has since acquired a negative charge may vary from 0 to some $\Delta\tau_{ex} \sim 10^{-6}$ s. The latter figure is a statistically *average* time of daemon-stimulated proton decay in the remainder of the captured nucleus. This may be any time for an actual event. Therefore, only about one half of the events of interest to us, which contain simultaneously both 2-6-keV and MeV-range scintillations, fall within the chosen time window. Significantly, such cases could be interpreted, in principle, as multiple-hits events too, because high-energy capture should excite scintillations in *several neighboring crystals*; also, the capture itself (not necessarily of the iodine but possibly of a Na nucleus as well) occurs with a high probability not in the crystal where the ion has been knocked out elastically. Multiple-hits events, however, do not exhibit annual periodicity, because, considered in terms of the WIMP hypotheses, multi-MeV scintillations do not carry useful information, with the result that the relevant events are disregarded in processing of experimental data.

## 5. Conclusions

We have to add to what has been said above that (*i*) only those 20-25% of the 2-6-keV iodine ions which have undergone channeling in the NaI(Tl) crystals are detected, whereas all the c-daemons entering a crystal (and even the material close by) are capable of generating MeV-range scintillations in the detector, including several neighboring crystals, and (*ii*) besides the SEECHO flux, there exist comparable-in-intensity NEACHO and GESCO (geocentric Earth-surface crossing orbit) fluxes of daemons producing recoil nuclei with $E_r < E_0 = 2$ keV and whose passage through matter is likewise accompanied by MeV-range capture of nuclei (these fluxes are modulated with $P = 0.5$ y, and their intensity appears to depend on latitude [6]). It thus follows that while the number of daemon-initiated MeV-range scintillations should exceed significantly that of the daemon-stimulated 2-6-keV events, about half of the latter should be accompanied, with the 600-ns window (and $\Delta\tau_{ex} \sim 10^{-6}$ s), by a MeV-range scintillation. These figures can be refined only by comprehensive calculations performed with due allowance for both experimental data and the crystal assembly geometry, as well as for interaction of daemons with the material (Cu, Pb, Cd) of the surrounding shielding.

Finally, one should not overlook the chance that the scenario of daemon interaction with matter described in Sec. 2 is incomplete and lacks some, possibly essential, aspects. This may eventually affect noticeably the strategy to be accepted in detection and investigation of daemons.

To mention just one point, it has already been said that the modes of the assumed daemon-stimulated proton decay are unclear. Indeed, it is conceivable that it is neutrons that are the first to decay in the daemon-containing nucleus, while protons, on reaching a certain excess number, fuse in a burst with a release of energy and ejection of positrons and neutrinos, as this should possibly occur in the vicinity of a negative daemon that has captured $Z$ free protons [18].

Similar processes can be envisaged to occur in attempts at using, say, organic low-atomic-weight scintillators for detection of daemons. One could conceive here of capture by the daemon of light nuclei into high (Rydberg) levels. If more than one nucleus have accumulated in these levels, they might fuse exothermally [4], again with the attendant ejection of scintillation-active particles.

Thus, the DAMA collaboration appears to have stumbled here on a new formidable challenge of switching from the WIMP to the daemon paradigm, a step readily apparent from their experimental data. In addition to the already reached, unprecedented high sensitivity of the system ($E_0 = 2$ keV), it would be desirable to envisage simultaneous (in a window of ~100-200 μs rather than 0.6 μs) detection of the scintillation events within an energy range extending over four orders of magnitude, i.e., up to ~10-20 MeV, a remarkable feat indeed, bearing in mind the inherent linearity of the characteristics of the scintillators and of the PM tubes. Also, the system should be triggered by a MeV-range event with $E \geq 3\text{-}5$ MeV, in which a negative c-daemon captures a new heavy nucleus in the scintillator (to ensure that it is indeed in the scintillator that the capture takes place, the scintillator assemblies should not apparently be isolated from one another by layers of material with a high atomic weight). Because a nucleus is knocked out elastically by a neutral c-daemon from the crystal lattice in $\Delta\tau_{ex} \sim 10^{-6}$ s, in order to be able to detect such precursors (in the 2-6-keV interval for the SEECHO c-daemons) and probable Na-capture events, one should provide also a possibility of recording events that have occurred, say, ≥5-10 μs before the system triggering (a standard procedure for present-day detector electronics, it is used in our experiments also [5,6,8,11,12]).

The proposed purposeful readjustment of the system based on the daemon paradigm would yield information not only on the properties of the SEECHO daemon flux (the intensity, velocity and direction of fall, etc.) but on streams from the NEACHOs, as well as from GESCOs into which the low-velocity NEACHO daemons are captured. Fine aspects of daemon interaction with matter would also be refined, including cross sections of capture and of elastic knocking out of various nuclei, the properties of charged and neutral c-daemons, the value of $\Delta\tau_{ex}$ and, possibly, the modes of daemon-stimulated nucleon decay. Hopefully, some other essential aspects of the daemon ideology itself would be refined.

To determine the trajectories of the daemons crossing the detector, it would be advisable to develop as the next step first not the 1-ton NaI system but rather a 0.5-ton cubic detector made up of 50 standard 9.7-kg NaI(Tl) crystals. In the neighboring successive layers, they should be arranged perpendicular to one another, thus making it possible to localize the track of the daemon in flight. The crystals could be combined pairwise by joining their end faces to ensure proper optical contact. Each such assembly, 50.8 cm long and 19.4 kg in weight, would be viewed from its outer ends by two its PM tubes working, as they are presently, in coincidence. Because the number of the PM tubes would not be changed and left as it is in LIBRA, the electronics likewise would not be doubled.

Looking forward optimistically into the future, it would certainly be desirable to develop and put in simultaneous operation several daemon-detection systems with overlapping parameters in different locations of the Earth, including the Southern hemisphere.